\documentclass[12pt,a4paper]{article}
\usepackage{amsmath}
\usepackage[dvips]{graphicx}
\input epsf
\usepackage{times}
\begin{document}
\begin{titlepage}
\title{Multiparticle production in the model with antishadowing}
\author{S. M. Troshin,
 N. E. Tyurin\\[1ex]
\small  \it Institute for High Energy Physics,\\
\small  \it Protvino, Moscow Region, 142280, Russia}
\normalsize
\date{}
\maketitle

\begin{abstract}
We discuss the role of absorbtion and antishadowing  in particle production.
We
reproduce power-like energy behavior of the mean multiplicity in the model
with antishadowing  and discuss  physical implications of such behavior for
the hadron structure.
 \\[2ex]
\end{abstract}
\end{titlepage}
\setcounter{page}{2}

\section*{Introduction}
Multiparticle production and  global  observables such as
mean multiplicity and its energy dependence alongside
with total, elastic and inelastic cross--sections  provide us  a clue
to the mechanisms of confinement and hadronization.
General principles  are very important in the nonperturbative sector of QCD and
unitarity which regulates the relative strength
of elastic and inelastic processes is the one of such principles.
There is no  universal, generally accepted
method to implement unitarity
in high energy scattering and as a result of this fact a related problem of the absorptive
corrections role and their sign
has a long history (cf. \cite{sachbla} and references therein).
However, a choice of particular unitarization scheme is not just a matter
of taste. Long time ago  arguments based on analytical properties of the scattering
amplitude  were put forward \cite{blan} in favor of the rational form of unitarization.
It was shown  that this form of unitarization reproduced
correct analytical properties of the scattering amplitude
in the complex energy plane much easier compare to the
exponential form.

Interest in unitarity limitations and the respective dependencies of the
global observables
was stimulated by the preparation
of the experimental program for the LHC and the future plans
to study soft interactions at the highest available energies.
For example, correct account for unitarity is  essential under theoretical
 estimates
of the Higgs production cross-section via  diffractive mechanisms.
The region of the LHC energies is the one where new, so called antishadow scattering
 mode can be observed. Such a mode naturally appears
 when the rational form of unitarization
 being exploited \cite{bdsphl}.  It has been demonstrated
 that this mode can be revealed at the LHC directly measuring
 $\sigma_{el}(s)$ and $\sigma_{tot}(s)$ \cite{reltot} (and not only by means of the
 analysis of impact parameter distributions).
 Antishadowing  leads to self--damping of the inelastic channels and
 asymptotically dominating role of elastic scattering, i. e.
 ${\sigma_{el}(s)}/{\sigma_{tot}(s)}\rightarrow 1$ at $s\to\infty$.
Immediate  question arises on consistency of this mechanism with
the growth of mean multiplicity in hadronic collisions with energy.
Moreover, many models and experimental data suggest a power-ilke energy dependence
of mean multiplicity\footnote{Recent discussions of power--like energy dependence
of the mean hadronic multiplicity and list of references to the older papers
can be found in \cite{menon}} and a priori the  compatibility
of such dependence  with  antishadowing is not evident.

In this note we apply
a rational ($U$--matrix)  unitarization
method \cite{umat}  to   consider  the global features
of multiparticle dynamics such as mean multiplicity and role
of absorptive  corrections. We show that it is possible to
reproduce power-like energy behavior of the mean multiplicity in the model
with antishadowing and discuss some physical implications.

\section{Multiparticle production in the $U$--matrix approach}
The rational form of unitarization
 is based on the relativistic generalization
 of the Heitler equation of radiation dumping \cite{umat}.
 In this approach the elastic scattering amplitude satisfies
unitarity since it is a solution  of  the
following equation  \begin{equation} F = U + iUDF
\label{xx} \end{equation}  presented here in the operator form.
 Eq.\ref{xx} allows  one to fulfill the unitarity provided the
 inequality \begin{equation} \mbox{Im} U(s,b) \geq 0 \end{equation}
is satisfied.
The form of the amplitude in the impact parameter representation is
\begin{equation}
f(s,b)=\frac{U(s,b)}{1-iU(s,b)}, \label{um}
\end{equation}
where $U(s,b)$ is the generalized reaction matrix. It is considered as an
input dynamical quantity similar to the eikonal function.
It is to be noted that
the analogous form for the scattering amplitude was obtained by Feynman in his
parton model for diffractive scattering which he has never published
(cf. \cite{ravn}).

In the impact parameter representation the unitarity equation
rewritten at high energies for the elastic amplitude $f(s,b)$ has the form
\begin{equation}
\mbox{Im} f(s,b)=|f(s,b)|^2+\eta(s,b) \label{unt}
\end{equation}
where the inelastic overlap function
\[
\eta(s,b)\equiv\frac{1}{4\pi}\frac{d\sigma_{inel}}{db^2}
\]
 is the sum of
all inelastic channel contributions.  It can be expressed as
a sum of $n$--particle production cross--sections at the
given impact parameter
\begin{equation}
\eta(s,b)=\sum_n\sigma_n(s,b),
\end{equation}
where \[\sigma_n(s,b)\equiv \frac{1}{4\pi}\frac{d\sigma_{n}}{db^2},\quad
\sigma_n(s)=8\pi\int_0^\infty bdb \sigma_n(s,b).\]
Inelastic overlap function
is related to $U(s,b)$ according to Eqs. (\ref{um}) and (\ref{unt}) as follows
\begin{equation}
\eta(s,b)=\frac{\mbox{Im} U(s,b)}{|1-iU(s,b)|^{2}}\label{uf}.
\end{equation}

Then the unitarity Eq. \ref{unt} points out that the elastic scattering
amplitude at given impact parameter value
is determined by the inelastic processes when the amplitude is
a pure imaginary one.
Eq. \ref{unt} imply the constraint
$|f(s,b)|\leq 1$
 while the ``black disk'' limit
 presumes inequality
$|f(s,b)|\leq 1/2$
and the elastic amplitude satisfying  the latter condition is a shadow of inelastic
processes. In its turn the imaginary
part of the generalized reaction matrix is the sum of inelastic channel
 contributions:
\begin{equation}
Im U(s,b)=\sum_n \bar{U}_n(s,b),\label{vvv}
\end{equation}
where $n$ runs over all the inelastic states and
\begin{equation}\label{gam}
\bar{U}_n(s,b)=\int d\Gamma_n |U_n(s,b,\{\xi_n\})|^2.
\end{equation}
In Eq. (\ref{gam}) $d\Gamma_n$ is the $n$--particle element of the phase space
volume.
The functions $U_n(s,b,\{\xi_n\})$ are determined by dynamics
 of $h_1+h_2\to X_n$ processes, where $\{\xi_n\}$ stands for the full
 set of respective kinematical variables.
  Thus, the quantity $\mbox{Im}U(s,b)$ itself
 is a shadow of the inelastic processes.
However, unitarity leads to  self--damping of the inelastic
channels \cite{bbl} in the sense that
 increase of the function $\mbox{Im}U(s,b)$ results in
decrease
 of the inelastic overlap function $\eta(s,b)$ when $\mbox{Im}U(s,b)$ exceeds unity
 (cf. Fig. 1).

\begin{figure}[hbt]
 \vspace*{-0.5cm}
 \begin{center}
 \epsfxsize=120  mm  \epsfbox{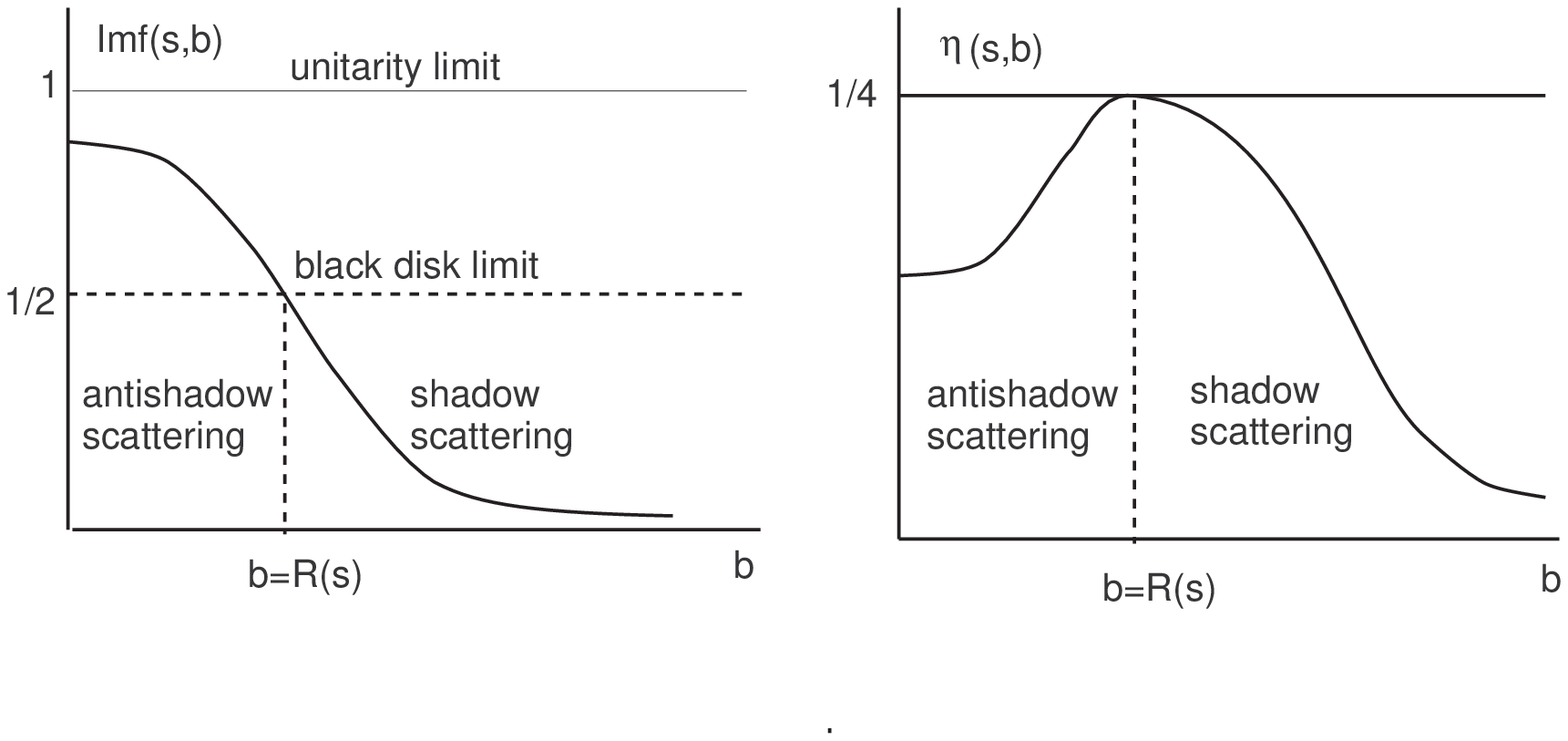}
 \end{center}
 \vspace{-1.5cm}
 \caption{Shadow and antishadow scattering regions}
 \end{figure}
Respective inclusive cross--section
\cite{tmf,gluas} which takes into account unitarity in the direct channel
  has the form \begin{equation}
\frac{d\sigma}{d\xi}= 8\pi\int_0^\infty
bdb\frac{I(s,b,\xi)} {|1-iU(s,b)|^2}.\label{un}
\end{equation}
The function $I(s,b,\xi)$ in Eq. (\ref{un}) is expressed via the functions
$U_n (s,b,\xi,\{\xi _{n-1}\})$  determined  by the dynamics
of the processes  $h_1+h_2\rightarrow
h_3+X_{n-1}$:
\begin{equation}\label{idef}
I(s,b,\xi)=\sum_{n\geq 3}n\int d\Gamma_n |U_n (s,b,\xi,\{\xi _{n-1}\})|^2
\end{equation}
and
\begin{equation}\label{sr}
\int I(s,b,\xi)d\xi =\bar n(s,b) \mbox{Im} U(s,b).
\end{equation}

 The kinematical variables $\xi$ ($x$ and $p_\perp$,
 for example) refer to the produced particle $h_3$ and
 the set of variables $\{\xi_{n-1}\}$ describe the system $X_{n-1}$
 of $n-1$ particles.

Now we turn to the mean multiplicity and consider first the corresponding
quantity in the impact parameter representation. As it follows from the above
the $n$--particle production
cross--section $\sigma_n(s,b)$
\begin{equation}\label{snb}
\sigma_n(s,b)=\frac{\bar{U}_n(s,b)}{|1-iU(s,b)|^2}
\end{equation}
Then the probability
\begin{equation}\label{pnb}
  P_n(s,b)\equiv\frac{\sigma_n(s,b)}{\sigma_{inel}(s,b)}=\frac{\bar{U}_n(s,b)}{\mbox{Im} U(s,b)}.
\end{equation}

Thus, we observe the cancellation of unitarity corrections in
the ratio of the cross-sections $\sigma_n(s,b)$ and $\sigma_{inel}(s,b)$.
Therefore the mean multiplicity in the impact parameter representation
\[
\bar n (s,b)=\sum_n nP_n(s,b)
\]
 is not affected by unitarity corrections  and
therefore cannot  be proportional
 to $\eta(s,b)$. This conclusion is  consistent with  Eq. (\ref{sr}).
The above mentioned  proportionality is a rather natural assumption in the framework
of the geometrical models, but it is in conflict with
the unitarization. Because of that the results  \cite{enk}
based on such assumption
 should  be taken with precaution.
However, the above cancellation of unitarity corrections
 does not take place for the quantity $\bar n (s)$ which we
 address  in the next section.

\section{Growth of mean multiplicity}

We use a   model for the hadron scattering
 described in \cite{csn}.
It is  based on the ideas of chiral quark models.
The picture of a hadron consisting of constituent quarks embedded
 into quark condensate implies that overlapping and interaction of
peripheral clouds   occur at the first stage of hadron interaction (Fig. 2).
Nonlinear field couplings  could transform then the kinetic energy to
internal energy and mechanism of such transformations was discussed
 by Heisenberg \cite{heis} and  Carruthers \cite{carr}.
As a result massive
virtual quarks appear in the overlapping region and  some effective
field is generated.
Valence constituent quarks  located in the central part of hadrons are
supposed to scatter simultaneously in a quasi-independent way by this effective
 field.

\begin{figure}[htb]
\hspace{2.5cm}
\epsfxsize=3in \epsfysize=1.85in\epsffile{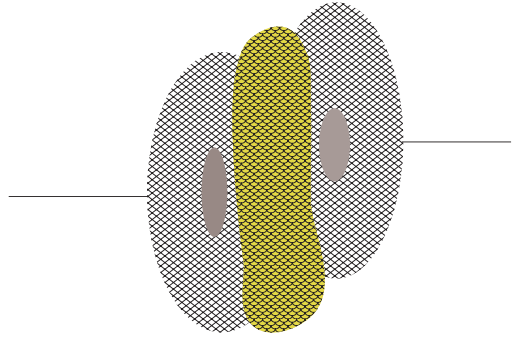}
 \caption[illyi]{Schematic view of initial stage of the hadron
 interaction.}
\label{ill5}
\end{figure}

 Massive virtual quarks play a role
of scatterers for the valence quarks in elastic scattering and
 their hadronization leads to
production of secondary particles in the central region.
 To estimate number
of such scatterers one could assume that  part of hadron energy carried by
the outer condensate clouds is being released in the overlap region
 to generate massive quarks. Then this number can be estimated  by:
 \begin{equation} \tilde{N}(s,b)\,\propto
\,\frac{(1-\langle k_Q\rangle)\sqrt{s}}{m_Q}\;D^{h_1}_c\otimes D^{h_2}_c
\equiv N_0(s)D_C(b),
\label{Nsbt}
\end{equation} where $m_Q$ -- constituent quark mass, $\langle k_Q\rangle $ --
average fraction of
hadron  energy carried  by  the constituent valence quarks. Function $D^h_c$
describes condensate distribution inside the hadron $h$, and $b$ is
an impact parameter of the colliding hadrons.

Thus, $\tilde{N}(s,b)$ quarks appear in addition to $N=n_{h_1}+n_{h_2}$
valence quarks. In elastic scattering those quarks are transient
ones: they are transformed back into the condensates of the final
hadrons. Calculation of elastic scattering amplitude has been performed
in \cite{csn}.
However,  valence quarks can excite a part of the cloud of the virtual massive
quarks and these virtual massive
 quarks will subsequently fragment into the multiparticle
final states. Such mechanism is responsible for the  particle production
in the fragmentation region and should lead to  strong correlations between
secondary particles. It means that correlations exist
between particles from the same (short--range correlations)
and different clusters (long--range correlations)
 and, in particular, the forward--backward
multiplicity correlations should be observed. This mechanism can be called
 as a correlated cluster
production mechanism. Evidently, similar mechanism should be significantly
reduced in $e^+e^-$--annihilation
processes and therefore large correlations are not to be expected there.

As it was already mentioned simple (not induced by interactions with valence
quarks) hadronization of massive $\tilde{N}(s,b)$
quarks leads to  formation of the multiparticle
final states, i.e. production of the secondary particles in the central region.
The latter should not provide any correlations in the multiplicity
distribution.

Remarkably,  existence of the massive quark-antiquark matter in the stage
preceding
hadronization seems to be
supported  by the experimental data obtained
at CERN SPS and RHIC (see \cite{biro} and references therein).

Since the quarks are constituent, it is natural to expect  direct
proportionality between a secondary particles multiplicity  and
number of virtual massive quarks appeared (due to  both mechanisms of multiparticle
production) in  collision of the  hadrons
with  given impact parameter:
\begin{equation}\label{mmult}
\bar n (s,b)=\alpha (n_{h_1}+n_{h_2})N_0(s)D_F(b)+ \beta N_0(s)D_C(b),
\end{equation}
with  constant factors $\alpha$ and $\beta$ and
\[
D_F(b)\equiv D_Q\otimes D_C,
\]
where the function $D_Q(b)$ is the probability amplitude of the interaction of
valence quark with the excitation of the effective field, which is in fact related
to the quark matter distribution in this hadron-like object called
the valence constituent quark \cite{csn}.
The mean multiplicity $\bar n(s)$ can be calculated according to the
formula
\begin{equation}\label{mm}
\bar n(s)= \frac{\int_0^\infty  \bar n (s,b)\eta(s,b)bdb}{\int_0^\infty \eta(s,b)bdb}.
\end{equation}
It is evident from Eq. (\ref{mm}) and Fig. 1 that the antishadow
mode with the peripheral profile of $\eta(s,b)$ suppresses the region of small
impact parameters the main contribution to the mean multiplicity is due to
 peripheral region of $b\sim R(s)$.

To make explicit calculations  we model for simplicity
the condensate distribution $D_C(b)$ and the impact parameter dependence
of the probability amplitude $D_Q(b)$
 of the interaction of
valence quark with the excitation of the effective field by the exponential forms,
and thus we use  exponential
 dependencies for the functions $D_F(b)$ and $D_C(b)$  with the
 different radii.
  Then the mean multiplicity
\begin{equation}\label{nsbex}
  \bar n (s,b)=\tilde\alpha N_0(s)\exp (-b/R_F)+\tilde\beta N_0(s)\exp (-b/R_C).
\end{equation}
The function $U(s,b)$   is
chosen as a product of the averaged quark amplitudes \begin{equation}
U(s,b) = \prod^{N}_{Q=1} \langle f_Q(s,b)\rangle \end{equation} in
accordance  since in the model valence quarks scatter in effective field simultaneously
and quasi-independently.
The $b$--dependence of  $\langle f_Q \rangle$ related to
 the quark formfactor $F_Q(q)$ has a simple form $\langle
f_Q\rangle\propto\exp(-m_Qb/\xi )$.
Thus, the generalized
reaction matrix (in a pure imaginary case) gets
the following  form \cite{csn}
\begin{equation} U(s,b) = ig\left [1+\alpha
\frac{\sqrt{s}}{m_Q}\right]^N \exp(-Mb/\xi ), \label{x}
\end{equation} where $M =\sum^N_{q=1}m_Q$.
At sufficiently high energies where increase of the total cross--sections
is quite prominent  we
can neglect the energy independent term  and rewrite
the expression for $U(s,b)$ as  \begin{equation}
U(s,b)=i{g}\left(s/m^2_Q\right)^{N/2}\exp (-Mb/\xi ).
\label{xh} \end{equation}

After calculation of the integrals (\ref{mm})
 we arrive to the power-like dependence
of the mean multiplicity $\bar n(s)$ at high energies
\begin{equation}\label{asm}
\bar n(s) = as^{\delta_F}+bs^{\delta_C},
\end{equation}
where
\[
\delta_{F}={\frac{1}{2}\left(1-\frac{\xi }{m_QR_{F}}\right)}\quad
\mbox{and}\quad \delta_{C}={\frac{1}{2}\left(1-\frac{\xi }{m_QR_{C}}\right)}.
\]
\begin{figure}[t]
\hspace{2cm}
\epsfxsize=3.5in \epsfysize=2.5in\epsffile{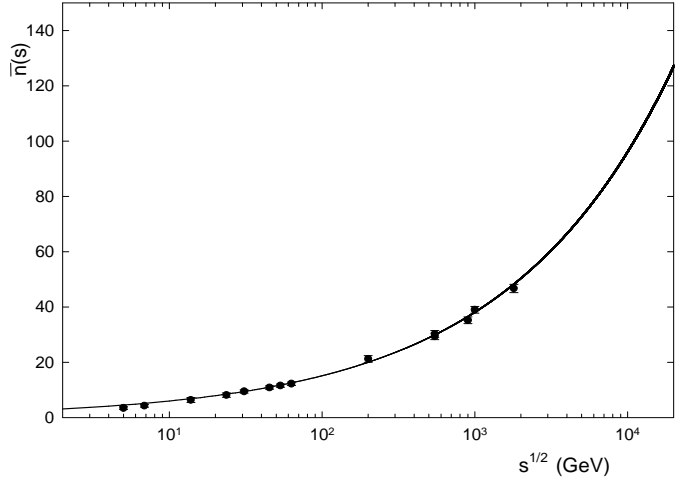}
 \caption[illyi]{Energy dependence of mean multiplicity, theoretical curve
  is given by the equation $\bar n(s)=as^\delta$ ($a=2.328$, $\delta = 0.201$); experimental
  data from the Refs. \cite{ua5}.}
\label{ill4}
\end{figure}
There are four free parameters in the model, $\tilde\alpha$, $\tilde\beta$ and
$R_F$, $R_C$, and the freedom
in their choice is translated to  $a$, $b$ and
 $\delta_F$, $\delta_C$.
 The value of  $\xi=2$ is fixed from the data on angular
 distributions \cite{csn} and for the mass of constituent quark
 the standard value $m_Q=0.35$ GeV was taken. However, fit to experimental data on the
 mean multiplicity leads to approximate equality $\delta_F\simeq \delta_C$ and
actually Eq. (\ref{asm}) is reduced to the two-parametric power-like energy dependence
 of mean multiplicity
\[
 \bar{n}=as^\delta,
 \]
 which is in good agreement with the experimental data (Fig. 3). Equality
 $\delta_F\simeq \delta_C$  means that variation of the correlation strength with
 energy is weaker than the power dependence and could be, e.g. a logarithmic one.
  From the comparison
 with the data  on mean multiplicity we obtain that
$\delta\simeq 0.2$, which corresponds to the effective masses, which are determined
by the respective radii ($M=1/R$),
$M_C\simeq M_F\simeq 0.3m_Q$, i.e. $M_F\simeq M_C\simeq m_\pi$.

The value of mean multiplicity expected at the LHC energy
($\sqrt{s}=14$ TeV) is about 110.
It is not surprising that it is impossible to differentiate contributions from the
two mechanisms of particle production at the level of mean multiplicity. The studies
of correlations are necessary for that purpose.
\section*{Conclusion}
It was shown that the model based on accounting unitarity \cite{csn} and extended to
multiparticle production provides a reasonable description of
the energy dependence of mean multiplicity leading to
its power-like growth with a small exponent. This result
is a combined effect of unitarity and existence of the phase preceding
 hadronization when massive quark--antiquark pairs are generated.
  It is worth noting again that power--like
energy dependence of mean multiplicity appears in various models and is
in a good agreement with  heavy--ion experimental data
too\footnote{Recent analysis of mean multiplicity with power--low growth
in Au+Au collisions at RHIC  is given in \cite{bars}}.

Multiplicity distribution $P_n(s,b)$ and mean multiplicity
$\bar n(s,b)$ in the impact parameter representation have
no absorptive corrections, but since
antishadowing leads to suppression of particle
production at small impact parameters and the main contribution to
the integral  multiplicity $\bar n(s)$ comes from
the region of $b\sim R(s)$. Of course, this prediction is to be valid for
the energy range where antishadow scattering mode starts to develop
(the quantitative analysis of the experimental data
 \cite{pras} gives the value: $\sqrt{s_0}\simeq 2$ TeV)
and is therefore consistent with the ``centrality'' dependence of the mean multiplicity
observed at RHIC \cite{phen}.

It is also worth  noting  that  no limitations follow from the general principles
 for the mean multiplicity, besides the well known one based
on the energy conservation law.
Having in mind relation (\ref{nsbex}), we could say  that the obtained power--like dependence
which takes into account unitarity effects could be considered as a kind of a saturated
upper bound  for the mean multiplicity growth.

\end{document}